\documentclass[11pt]{article}
\pdfoutput=1
\usepackage{jheppub}
\usepackage{bbm}

\title{Constraints on the NMSSM from the oblique parameters}

\preprint{BI-TP 2012/16}

\author{M. Maniatis}
\author{and Y. Schr\"oder}

\affiliation{Fakult\"at f\"ur Physik, Universit\"at Bielefeld,
33615 Bielefeld, Germany}

\emailAdd{maniatis@physik.uni-bielefeld.de}
\emailAdd{yorks@physik.uni-bielefeld.de}

\abstract{Electroweak precision measurements, 
encoded in the oblique parameters,
give strong constraints
on physics beyond the Standard Model.
The oblique parameters~$S$, $T$, $U$ ($V$, $W$, $X$)
are calculated in the next-to-minimal supersymmetric model~(NMSSM).
We outline the calculation of the 
oblique parameters in
terms of one-loop gauge-boson selfenergies and
find sensitive restrictions
for the NMSSM parameter space.}

\begin{document}

\maketitle
\flushbottom

\section{Introduction}
\label{se:intro}

The precision measurements of 
the electroweak parameters give
stringent constraints on physics
beyond the Standard Model~(SM).
A very elegant method to
systematically confront
the electroweak precision measurements
with new physics
is given by the oblique
parameters~$S$, $T$, $U$~\cite{Peskin:1990zt,Kennedy:1990ib,Peskin:1991sw}.
These three parameters allow
to restrict any physics 
beyond the SM, under the following three conditions:
\begin{itemize}
\item The physics beyond the SM
has to obey $SU(2)_L \otimes U(1)_Y$
gauge symmetry, that is,
there are no additional electroweak gauge
bosons compared to the SM.
\item The couplings
of new particles to light fermions
have to be suppressed. That is,
the main contribution
of couplings beyond the SM
to four--fermion scattering 
originates from
the change in the
self-energies of
the gauge-boson propagators.
These contributions
are called {\em oblique} corrections.
The suppressed contributions which
may for instance appear in box diagrams with
four external fermions or in vertex corrections are
called {\em non-oblique} corrections.
\item New physics enters only at
a scale large compared to the electroweak scale.
\end{itemize}
From the second condition it is clear that the
oblique parameters are expressed in terms of
gauge-boson self-energies, as was shown in detail in 
Ref.~\cite{Peskin:1991sw}. The main argument is that 
the electroweak precision measurements probe weak-interaction processes
with light external fermions of mass $m_f$
(at cms energies on the electroweak scale),
wherein vertex- and box-type correction are suppressed by factors
of $m_f^2/m_Z^2$ as compared to the self-energy loop corrections. 

However, many models beyond the SM are expected
to have effects at a scale not too far from the
electroweak scale, which is given by the
vacuum--expectation value of the 
neutral SM Higgs boson component~$v_0\approx 174$~GeV.
In order to weaken the third condition for
the oblique parameters~$S$, $T$, $U$ above,
allowing new physics to enter already at
a scale somewhat
larger than the electroweak scale, the
oblique parameters were extended to the
six parameters~$S$, $T$, $U$, $V$, $W$, $X$~\cite{Maksymyk:1993zm,Burgess:1993mg}.
The explicit expressions for these
oblique parameters read
\begin{eqnarray}
\label{eq:oblique}
S &=& \frac{4 s_W^2 c_W^2}{\alpha} \bigg[
\frac{\Pi_{ZZ}(m_Z^2)-\Pi_{ZZ}(0)}{m_Z^2} 
- \frac{c_W^2 -s_W^2}{s_W c_W} \Pi'_{Z\gamma}(0) - \Pi'_{\gamma\gamma}(0)
\bigg]\,,\nonumber\\
T &=& \frac{1}{\alpha} \bigg[
\frac{\Pi_{WW}(0)}{m_W^2} - \frac{\Pi_{ZZ}(0)}{m_Z^2}
\bigg]\,,\nonumber\\
U &=& \frac{4 s_W^2}{\alpha} \bigg[
\frac{\Pi_{WW}(m_W^2)-\Pi_{WW}(0)}{m_W^2} 
- c_W^2 
\frac{\Pi_{ZZ}(m_Z^2)-\Pi_{ZZ}(0)}{m_Z^2} 
- 2 s_W c_W
\Pi'_{Z\gamma}(0) 
- s_W^2
\Pi'_{\gamma\gamma}(0)
\bigg]\,,\nonumber\\
V &=& \frac{1}{\alpha} \bigg[
\Pi'_{ZZ}(m_Z^2) 
- \frac{\Pi_{ZZ}(m_Z^2)-\Pi_{ZZ}(0)}{m_Z^2} 
\bigg]\,,\nonumber\\
W &=& \frac{1}{\alpha} \bigg[
\Pi'_{WW}(m_W^2) 
- \frac{\Pi_{WW}(m_W^2)-\Pi_{WW}(0)}{m_W^2} 
\bigg]\,,\nonumber\\
X &=& -\frac{s_W c_W}{\alpha} \bigg[
\frac{\Pi_{Z\gamma}(m_Z^2)}{m_Z^2} - \Pi'_{Z\gamma}(0)
\bigg]\,.
\end{eqnarray}
The quantities $\Pi_{G_1 G_2}(s)$ 
with $G_{1/2} \in \{\gamma, W, Z\}$ 
denote the new
contributions to the transverse part of the 
self-energies at a momentum-squared scale~$s$ -- compared to
the SM, 
\begin{equation}
\label{eq:Pi}
\Pi_{G_1 G_2}(s) = \Pi_{G_1 G_2}^{\text{new}}(s)-\Pi_{G_1
  G_2}^{\text{SM}}(s)\,.
\end{equation}
The derivatives of the self-energies $\Pi_{G_1 G_2}(s)$
with respect to the scale~$s$ are denoted by 
$\Pi'_{G_1 G_2}(s_0)={\rm d}\Pi_{G_1 G_2}(s)/{\rm d}s|_{s=s_0}$. 
The fact that only relatively few parameters 
(besides $\Pi(s)$ for $s\in\{0,m_W^2,m_Z^2\}$ only
$\Pi'(s)$ at the same low-energy scales)
enter in Eq.~\eqref{eq:oblique}
reflects the observation that precision measurements are
made only by two-particle scatterings on light fermions 
at those few scales,
as explained in detail in Ref.~\cite{Maksymyk:1993zm}.
Finally, $s_W=\sin(\theta_W)$ and
$c_W=\cos(\theta_W)$ contain the usual weak Weinberg
mixing angle~$\theta_W$, and $\alpha$ denotes the fine-structure constant.

Having defined the oblique parameters, electroweak precision observables, like
for instance the $W^\pm$-boson mass,
may be expressed in terms of these parameters.
Constraints on the oblique parameters are gained 
via a global fit to the electroweak precision measurements; see
e.g. Ref.~\cite{Nakamura:2010zzi}.
Being exactly zero within the SM, these global fits result in error bands
for the six parameters of Eqs.~\eqref{eq:oblique}, \
see Eq.~\eqref{eq:obliqueexp} below, hence potentially constraining 
the size of effects from new physics.

In this paper we compute the oblique
parameters~$S$, $T$, $U$, $V$, $W$, $X$ 
of Eq.~\eqref{eq:oblique} in the next-to minimal supersymmetric
extension of the SM~(NMSSM); for reviews
of the NMSSM, we refer to Refs.~\cite{Maniatis:2009re,Ellwanger:2009dp}. 
The NMSSM receives a lot of attention in recently
-- in particular, it possesses a scale invariant
superpotential, a much richer
Higgs sector, and a fifth neutralino
compared to the minimal
supersymmetric extension~(MSSM).
Noting that in the fermion--fermion
interactions there appear in principle also 
non-oblique corrections in the NMSSM,
here we assume that the non-oblique
corrections are negligible.

There are several computations of 
electroweak precision observables in the NMSSM.
Let us mention the study of
the $Z^0$ boson width~\cite{2001yb}
as well as the study of the $W^\pm$-mass
and the $Z$ boson decay into leptons~\cite{Domingo:2011uf}.
Let us remark that there exist similar approaches
for the case of the 
MSSM \cite{Dedes:1998hg,Heinemeyer:2004gx,RamseyMusolf:2006vr,Cho:2011rk}.
Since the parameter space of the NMSSM is very large,
there are different approaches to phenomenological
studies of this model. 
In Refs.~\cite{Ellwanger:2006rn,Djouadi:2008uw,Gunion:2012zd}, 
for instance, the constrained version of the
NMSSM is considered where it is assumed that various
masses and couplings unify at the GUT scale.
Another approach is to consider specific benchmarks scenarios,
representing different regimes in parameter 
space \cite{AbdusSalam:2011fc,King:2012is}.
In our numerical examples below, we shall adopt the former approach.

\section{Details of the calculation}

For the prediction of the oblique parameters of Eq.~\eqref{eq:oblique}
we need to compute the transverse parts of
the one-loop self-energies
$\Pi_{G_1 G_2}(s) = \Pi_{G_1 G_2}^{\text{NMSSM}}(s)-\Pi_{G_1 G_2}^{\text{SM}}(s)$,
where $G_1$, $G_2$ denote the gauge bosons $\gamma$, $W^\pm$, $Z^0$.
The self-energies 
with exclusively leptons, quarks, and gauge bosons in the
loops are exactly the same in 
$\Pi_{G_1 G_2}^{\text{NMSSM}}(s)$ and
$\Pi_{G_1 G_2}^{\text{SM}}(s)$ and
therefore do not need to be evaluated.
As a consistency check, however, we confirmed
this analytically.

\begin{table}[t]
\begin{center}
\begin{tabular}{cccccccc}
set\#&
$M_0^{\text{GUT}} $ &
$M_{\text{1/2}}^{\text{GUT}} $ &
$A_0^{\text{GUT}} $ &
$A_\kappa^{\text{GUT}} $ &
$\tan(\beta)^{\text{MSUSY}}$ &
${\rm sgn}(\mu)$ &
$\lambda^{\text{MSUSY}}$
\\
\hline
1 & 500 & 500 & $-800$ & $-100$ & $5$ & $+$ & $0.15$\\
2 & 500 & 500 & $-800$ & $-1500$ & $1.7$ & $+$ & $0.5$\\
3 & 100 & 200 & $-700$ & $-75$ & $5$ & $+$ & $0.2$\\
\end{tabular}
\end{center}
\caption{\label{tab:parameters}
Parameter values for the example studies in case of the constrained NMSSM
($M_i$ and $A_i$ in GeV).
These sets are inspired by the ranges given in Figs.~1--3 
of Ref.~\cite{Ellwanger:2006rn}.
}
\end{table}

Since the Higgs sector of the NMSSM
is not a simple extension of the SM Higgs
sector, we have to consider in
$\Pi_{G_1 G_2}^{\text{SM}}(s)$
all contributions which contain
the SM Higgs boson~$H_{\text{SM}}$.
In the self-energies of the
NMSSM we have to consider all contributions
which involve scalar neutrinos~$\tilde{\nu}$,
scalar leptons~$\tilde{l}$, scalar up- and down-
type quarks~$\tilde{u}$, $\tilde{d}$,
neutralinos~$\chi^0$, charginos~$\chi^+$,
the neutral Higgs bosons~$H_1$, $H_2$, $H_3$, $A_1$, $A_2$,
the pair of charged Higgs bosons~$H^\pm$ and
the Goldstone bosons~$G^0$, $G^+$.
All Feynman diagrams of the self-energy contributions
to the oblique
parameters are shown in App.~\ref{se:appA}.
Let us note that we consider the 
most general NMSSM in our computation.
In particular we allow for
CP violation in the Higgs sector,
such that the neutral Higgs bosons
$H_i/A_j$
are not necessarily CP even/odd,
respectively; for details
see for instance Ref.~\cite{Maniatis:2009re}.

\begin{figure}
\centerline{\includegraphics[width=0.9\textwidth]{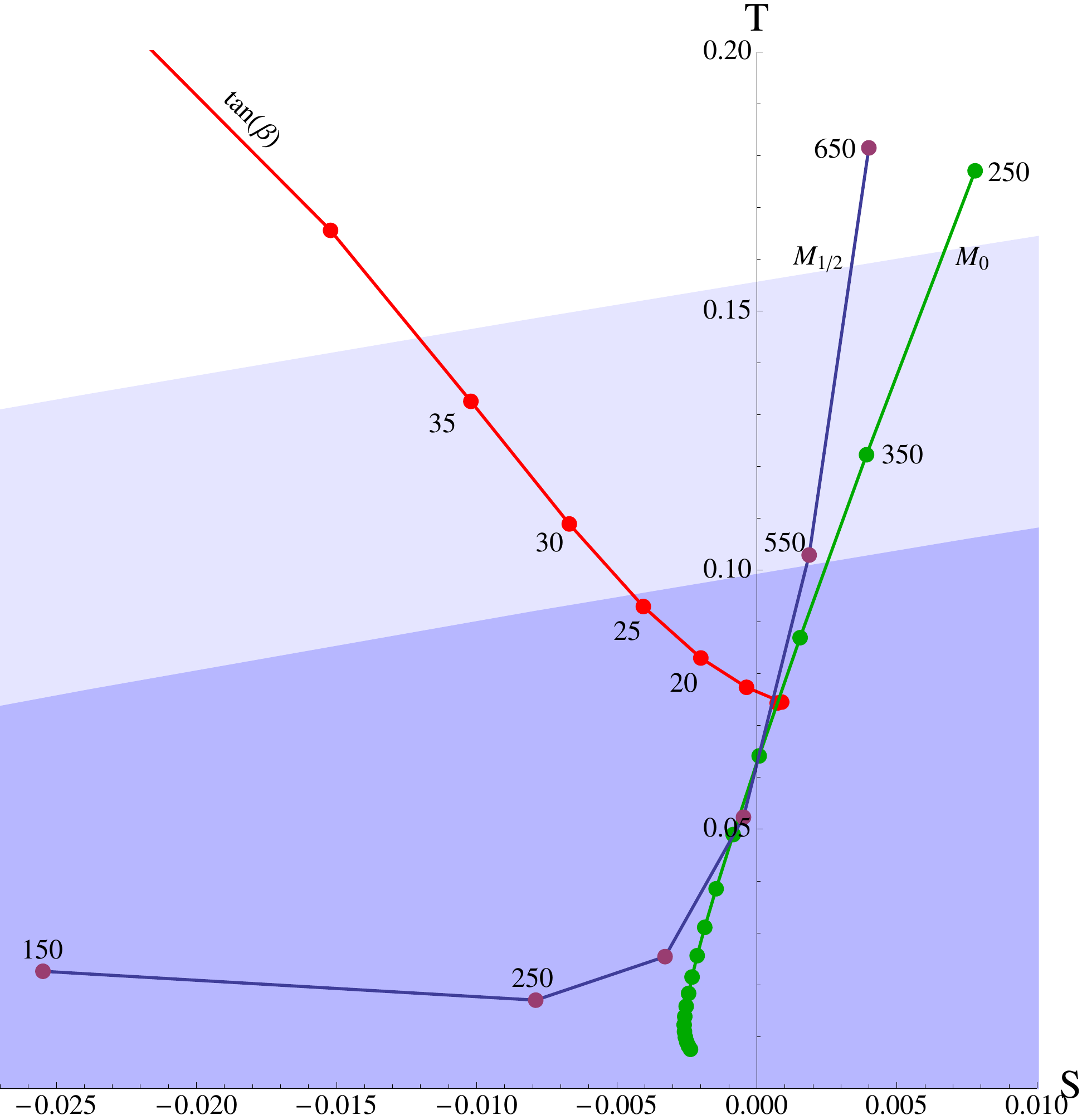}}
\caption{\label{fig:plotST1}
Oblique parameters~$S$ and $T$ in the NMSSM
in the constrained case with parameters 
$\tan(\beta) = 5$,
$M_0 = M_{1/2} = 500$~GeV,
$A_0 =-800$~GeV,
$A_\kappa= -100$~GeV,
${\rm sgn}(\mu)=+$, and
$\lambda=0.15$ 
(first row of Tab.~\ref{tab:parameters})
with
the parameters $\tan(\beta)$,
$M_0$, 
$M_{\text{1/2}}$ varied as
indicated in the figure.
The shaded
regions show
the $1\sigma$ and $2\sigma$ error ellipses of the electroweak
precision measurements fitted to $S$ and $T$
corresponding to Eq.~\eqref{eq:obliqueexp}~\cite{Nakamura:2010zzi}.
}
\end{figure}

The NMSSM Feynman rules are implemented 
in the {\tt FeynRules} program package~\cite{Christensen:2008py,Duhr:2011se}
following the conventions of Ref.~\cite{arXiv:0801.0045}.
As a caveat, let us remark here that the Goldstone components 
of the neutral Higgs boson squared mixing matrix have to 
be carefully constructed such as to guarantee unitarity,
which is violated by the parameters chosen in the model file 
{\tt nmssm.fr}.
We link this list of Feynman rules with the packages 
{\tt FeynArts}/{\tt FormCalc}~\cite{Hahn:2000kx},
resulting in analytic expressions for the various one-loop self-energies
in terms of basic scalar master integrals.
Next, we assemble the parameters of Eq.~\eqref{eq:oblique}
and numerically evaluate the results using
the program package {\tt LoopTools}~\cite{Hahn:1998yk}.
We observe that all ultraviolet singularities
cancel between the different
self-energies in the oblique parameters.
On a more technical note, 
the matrix $\gamma^5$ is treated naively
(that is, anticommuting)
with $(\gamma^5)^2=\mathbbm{1}_{4\!\times\!4}$,
while we have checked explicit gauge parameter independence.

\section{Results}
\label{se:results}

As a simple numerical example we
assume unification of
all scalar masses~$M_0$, fermion masses~$M_{1/2}$, 
and trilinear couplings~$A_0$ (except for~$A_\kappa$,
which is considered separately) at the GUT scale. 
This scenario is usually called constrained NMSSM (cNMSSM). 
Furthermore, the ratio of the vacuum-expectation value
of the two Higgs-boson doublets, $\tan(\beta)$,
the Higgs coupling parameter~$\lambda$, and
the sign of~$\mu$ have to be fixed in addition to
the parameters of the SM. 
The computation of the mass spectra and
mixing angles at the electroweak scale 
is performed with the program package~{\tt NMSPEC}~\cite{Ellwanger:2006rn}.
Let us note that our calculation of the
oblique parameters is
performed in the general NMSSM such that
the oblique parameters for arbitrary 
parameter values can be easily computed.
The program code for the oblique parameters
is available as C-code from the URL~\cite{ccode}.

The explicit values for the NMSSM parameters
we choose in our numerical examples are given in Table~\ref{tab:parameters}.
These parameter sets
are inferred from the
figures presented in Ref.~\cite{Ellwanger:2006rn}.
The scales at which the NMSSM parameters are 
fixed are written as a superscript,
with MSUSY and GUT
the supersymmetry breaking scale, respectively, the
grand unification scale -- both scales are 
derived from the input parameters in~{\tt NMSPEC}.

\begin{figure}
\centerline{\includegraphics[width=0.45\textwidth]{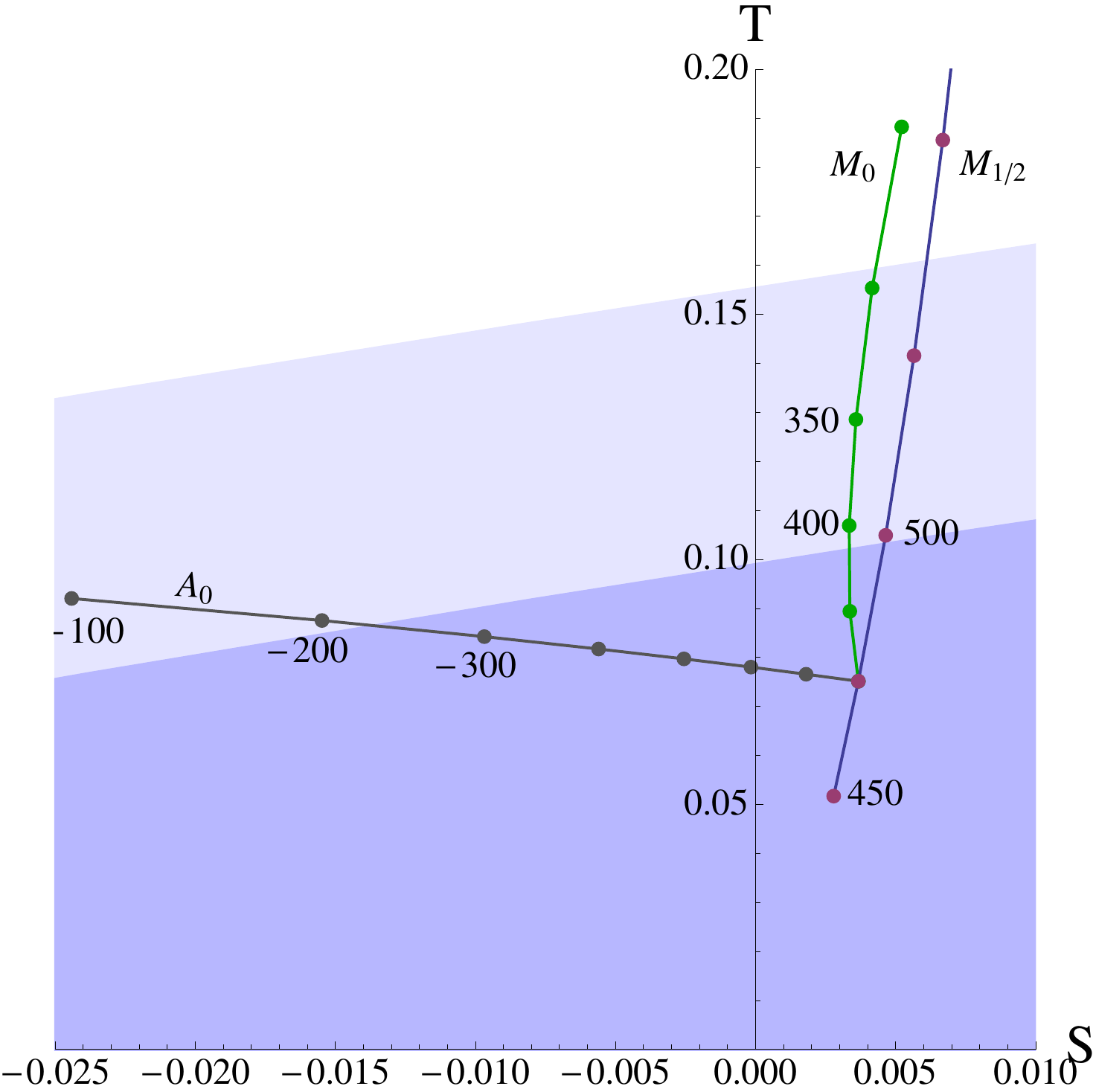}
\hspace*{8mm}
\includegraphics[width=0.45\textwidth]{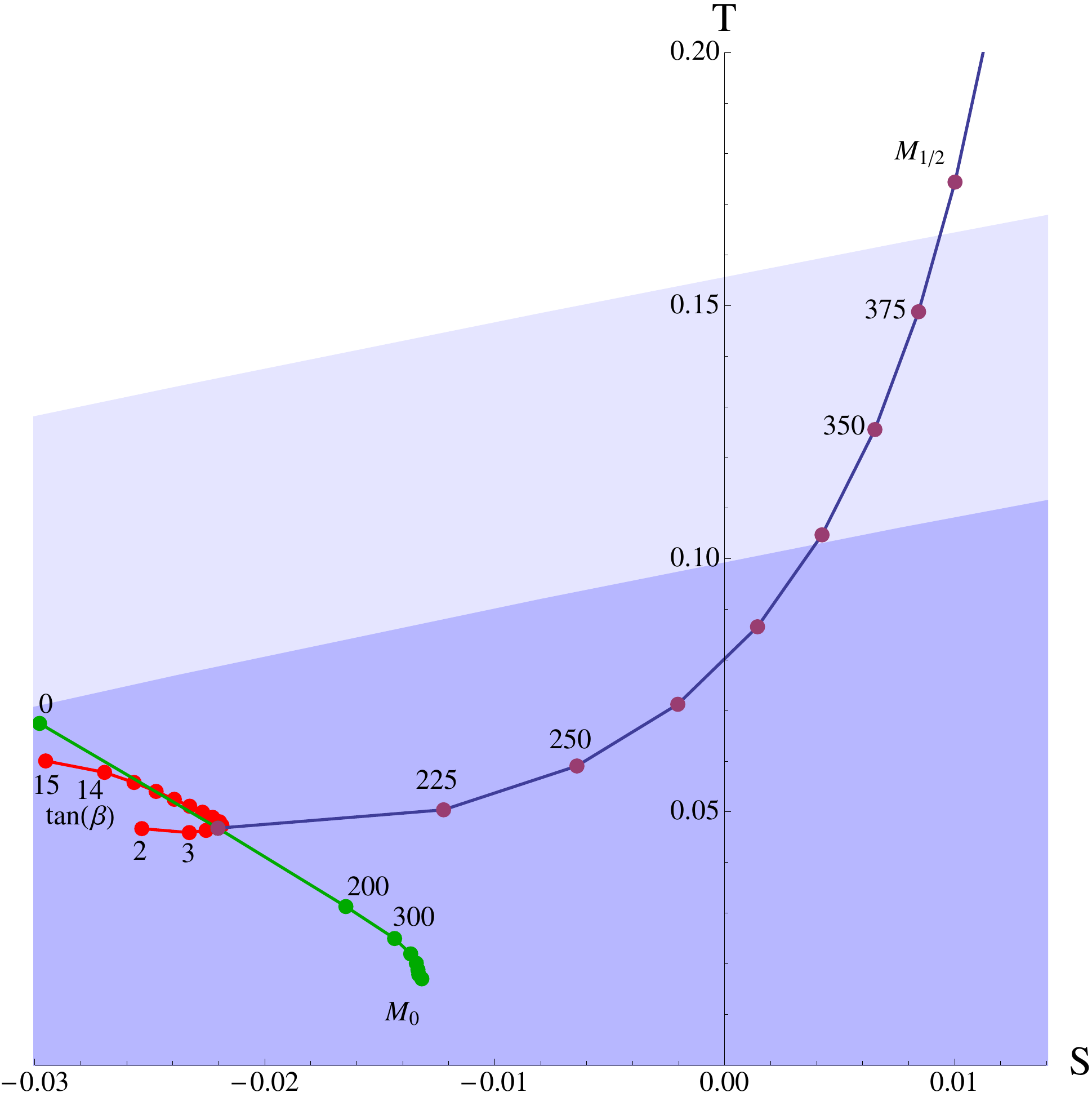}}
\caption{\label{fig:plotST2}
Left: Oblique parameters~$S$ and $T$ in the NMSSM in
the constrained case with  
central parameters
$\tan(\beta) = 1.7$,
$M_0 = M_{1/2} = 500$~GeV,
$A_0=-800$~GeV,
$A_\kappa= -1500$~GeV,
${\rm sgn}(\mu)=+$, and
$\lambda=0.5$ (second row in Tab.~\ref{tab:parameters}) with
variation of parameters
as indicated in the figure.
Right: Same as left but with central parameters
$\tan(\beta) = 5$,
$M_0 = 100$~GeV,
$M_{\text{1/2}}=200$~GeV,
$A_0=-700$~GeV,
$A_\kappa= -75$~GeV,
${\rm sgn}(\mu)=+$, and
$\lambda=0.2$.}
\end{figure}

In Figs.~\ref{fig:plotST1}, \ref{fig:plotST2} we present the
results for the oblique parameters~$S$ and $T$
for the different parameter sets given in Table~\ref{tab:parameters}.
All other oblique parameters turn out to be rather small
and are therefore not shown explicitly. 
In the Figures we vary
successively 
the parameters $A_0$, $\tan(\beta)$, $M_0$, and $M_{\text{1/2}}$ 
about the
central values from~Table~\ref{tab:parameters} as 
indicated in the figures (where we suppress the superscripts MSUSY and GUT).
From the lines
we see
how the oblique parameters~$S$ and $T$ change under variations of the
parameter values.
We also draw the $1\sigma$ and $2\sigma$ error ellipse corresponding
to the recent experimental fits to
$S$ and $T$~\cite{Nakamura:2010zzi}:
\begin{equation}
\label{eq:obliqueexp}
S=0.01\pm 0.1, \quad
T=0.03\pm 0.11, \quad
\rho = 0.87.
\end{equation}
Here, $\rho$ denotes the correlation coefficient.
Note that in this fit a SM Higgs-boson
mass of~$m_{H_\text{SM}}=117$~GeV is assumed,
which we also use consistently as a parameter value in 
the SM self-energies.

As expected, in our numerical examples we find suppressed contributions
to the oblique parameters~$V$, $W$, $X$, which is due to
the large masses 
of the additional particles as compared
to the electroweak scale. From the sensitivity
of the oblique parameters under variations of NMSSM parameters 
is clearly visible:
for the central parameter set~1 in Table~\ref{tab:parameters}
we infer from~Fig.~\ref{fig:plotST1}, that the $2\sigma$ error ellipse 
constrains $\tan(\beta)\lesssim 40$, $M_0 \gtrsim 250$~GeV, 
and $M_{1/2}\lesssim 650$~GeV. For the other central values in
Table~\ref{tab:parameters} we can easily read off the constraints
from Fig.~\ref{fig:plotST2}.

\section{Conclusions}
\label{se:con}

For a large class of models beyond the Standard Model,
the so-called oblique parameters give 
very sensitive constraints coming
from electroweak precision measurements. We have
computed the set of extended oblique parameters~$S$, $T$, $U$, $V$, $W$, $X$
for the next-to-minimal supersymmetric model~(NMSSM).

We have presented numerical examples
with the parameters of the NMSSM
chosen in a constrained case, as explained in Sec.~\ref{se:results}. 
We observe the oblique parameters $S$ and $T$ to be highly sensitive
on variations of the model parameters.
In fact, fairly modest changes of the NMSSM parameters
easily violate the constraints from the electroweak
precision measurements. 

The oblique parameters
have been computed for the general case, 
in particular with a general CP violating Higgs sector, such that they
may be applied to arbitrary parameter values,
in a more complete parameter scan, which we reserve for future work.

\acknowledgments

We are grateful to B.~Fuks for quick response concerning 
an update of the NMSSM model file contained in {\tt FeynRules}.
The work of Y.S. is supported by the Heisenberg program of the Deutsche
Forschungsgemeinschaft (DFG), contract no. SCHR 993/1.

\appendix
\section{Feynman diagrams for the oblique parameters}
\label{se:appA}

Here we present the Feynman diagrams which contribute to the
oblique parameters of  Eq.~\eqref{eq:oblique}.
For self-energy diagrams which exclusively
have leptons, quarks, and gauge
bosons in the loops, the contributions
to 
$\Pi_{G_1 G_2}^{\text{NMSSM}}(s)$ and
$\Pi_{G_1 G_2}^{\text{SM}}(s)$ exactly
cancel in Eq.~\eqref{eq:Pi} and do not have to be computed.

\begin{figure}[t]
\centerline{\includegraphics[width=0.45\textwidth]{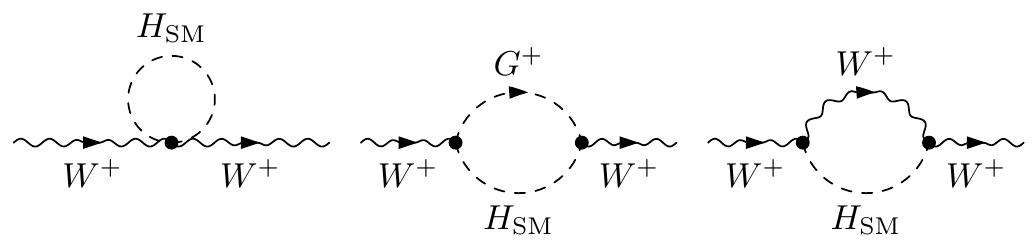} \qquad
\includegraphics[width=0.45\textwidth]{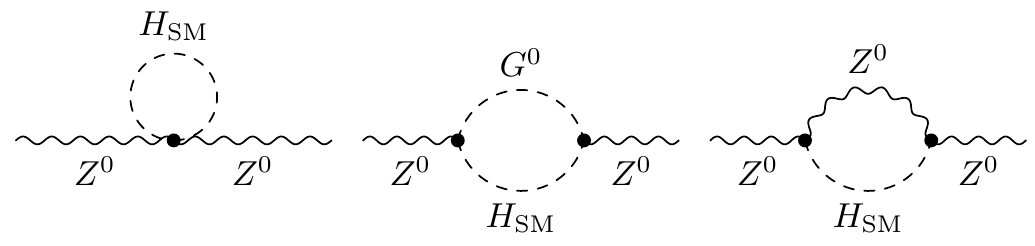}}
\caption{\label{fig:SM_self}
Feynman diagrams for the self-energies 
$\Pi_{W W}^{\text{SM}}(s)$
and
$\Pi_{Z Z}^{\text{SM}}(s)$
which contribute to the oblique
parameters. All other diagrams
vanish, respectively, cancel
with the corresponding diagrams in
$\Pi_{W W}^{\text{NMSSM}}(s)$
and
$\Pi_{Z Z}^{\text{NMSSM}}(s)$.
}
\end{figure}

The contributions to
$\Pi_{G_1 G_2}^{\text{SM}}(s)$
consist of diagrams which
contain the SM Higgs boson~($H_\text{SM}$) in the loop.
There are only contributions 
of this kind to the 
$W^+$ and $Z^0$ self-energies
as shown in Fig.~\ref{fig:SM_self}.

\begin{figure}[t]
\centerline{\includegraphics[width=0.95\textwidth]{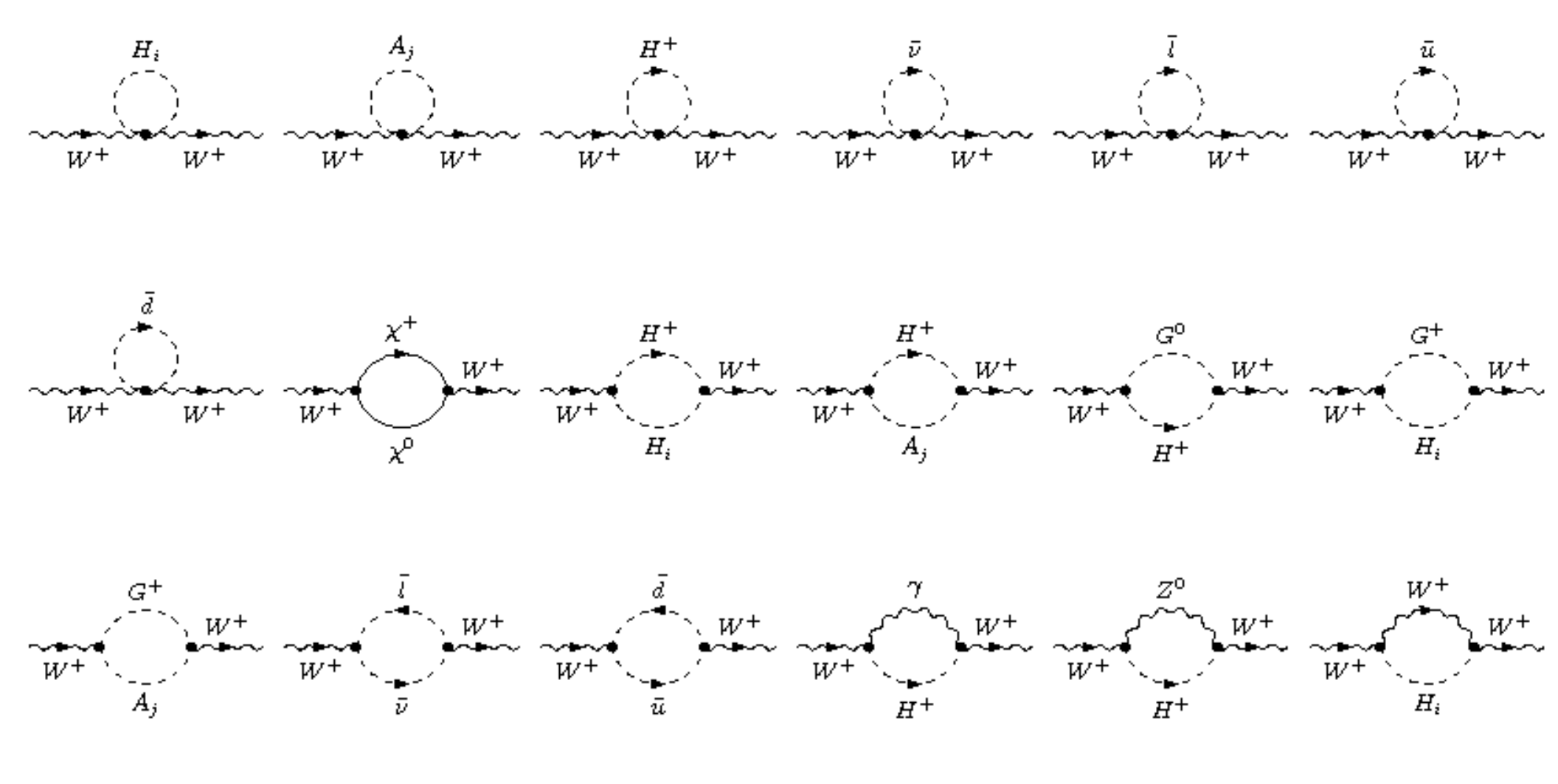}}
\caption{\label{fig:NMSSM_W}
Feynman diagram contribution to the
self-energy $\Pi_{W W}^{\text{NMSSM}}(s)$ }
\end{figure}
\begin{figure}[t]
\centerline{\includegraphics[width=0.9\textwidth]{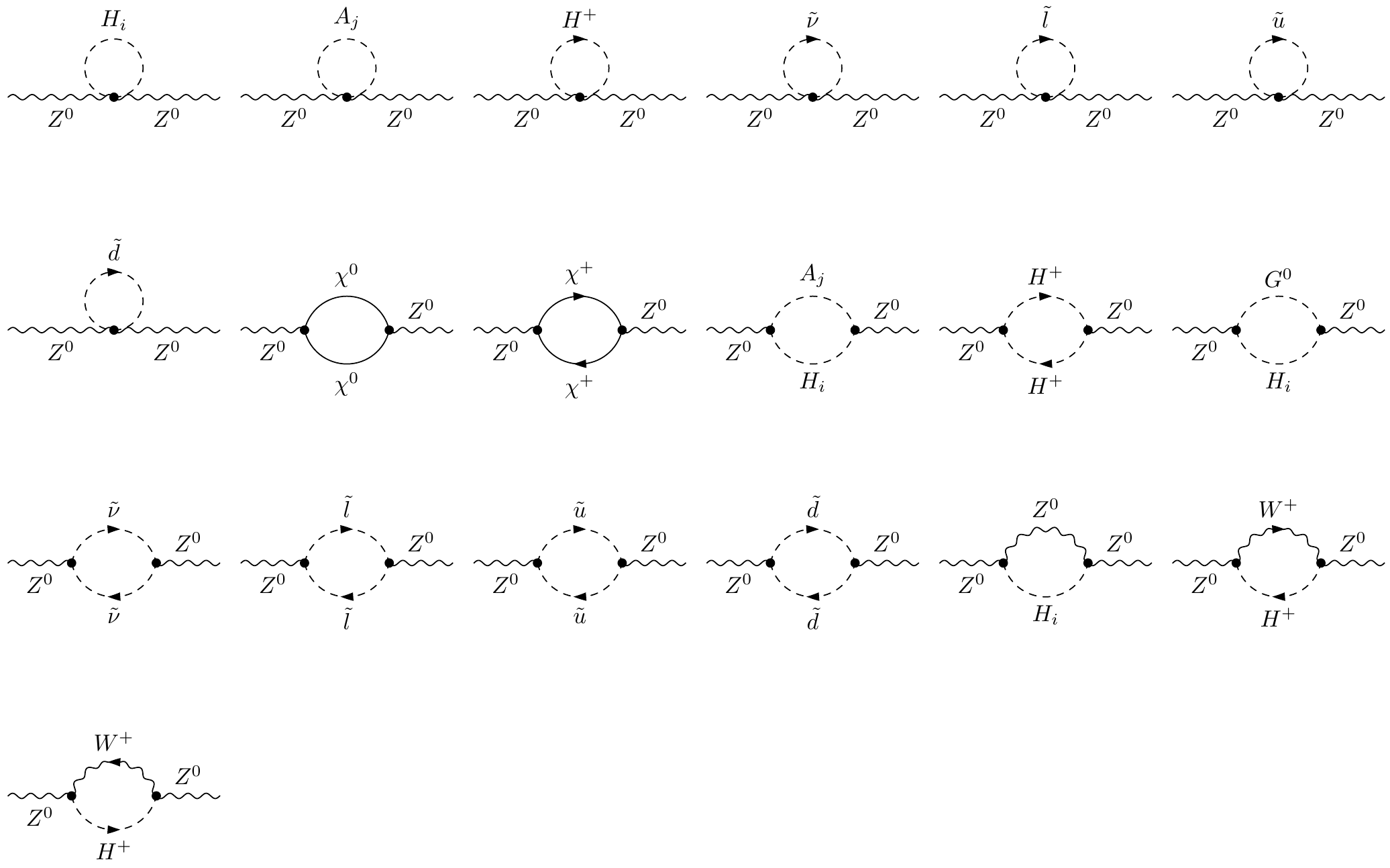}}
\caption{\label{fig:NMSSM_Z}
Feynman diagram contribution to the
self-energy $\Pi_{Z Z}^{\text{NMSSM}}(s)$ }
\end{figure}
\begin{figure}[t]
\centerline{\includegraphics[width=0.9\textwidth]{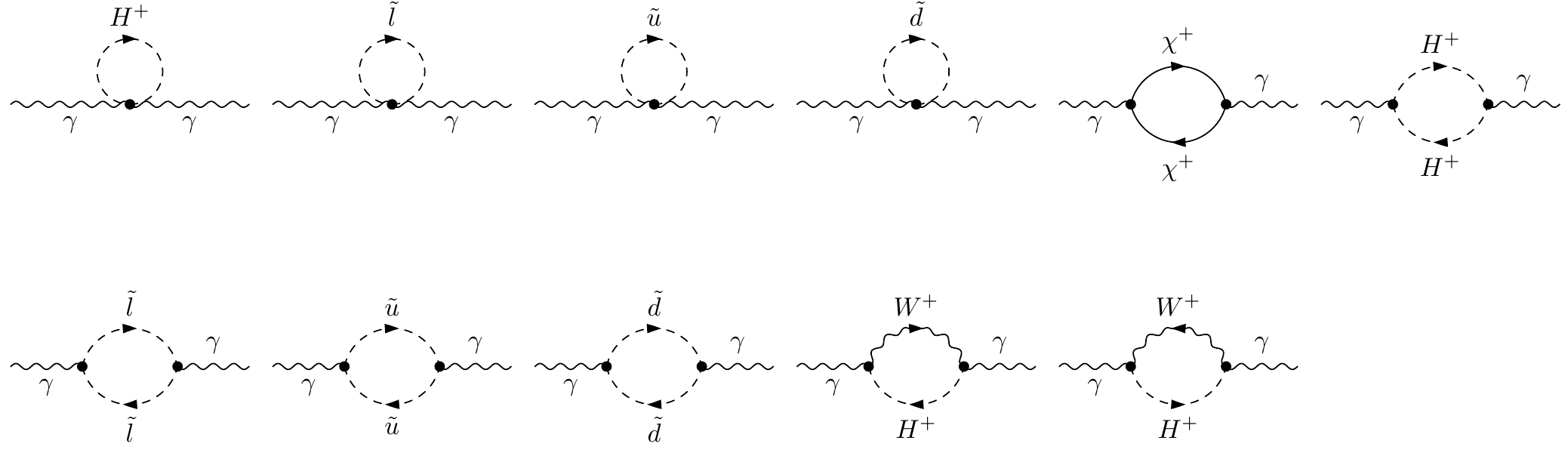}}
\caption{\label{fig:NMSSM_g}
Feynman diagram contribution to the
self-energy $\Pi_{\gamma \gamma}^{\text{NMSSM}}(s)$ }
\end{figure}
\begin{figure}[t]
\centerline{\includegraphics[width=0.9\textwidth]{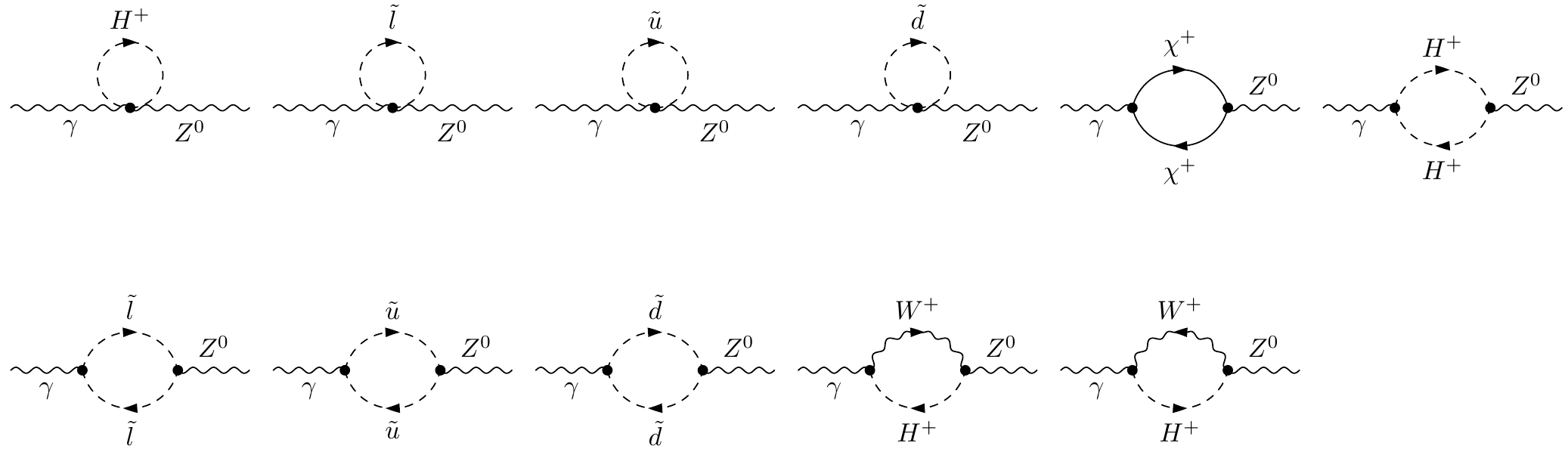}}
\caption{\label{fig:NMSSM_Zg}
Feynman diagram contribution to the
self-energy $\Pi_{Z \gamma}^{\text{NMSSM}}(s)$ }
\end{figure}
We also show all self-energy diagrams 
contributing to the NMSSM part
of the oblique parameters.
These diagrams involve scalar neutrinos~$\tilde{\nu}$,
scalar leptons~$\tilde{l}$, scalar up- and down-
type quarks~$\tilde{u}$, $\tilde{d}$,
neutralinos~$\chi^0$, charginos~$\chi^+$,
as well as the neutral Higgs bosons~$H_i$, $A_j$,
the charged Higgs bosons~$H^\pm$ as well as
the Goldstone bosons~$G^0$, $G^+$. All
other contributions, for instance
the self-energy with a lepton loop, cancel with
the corresponding SM contribution.

The $W^+$, $Z^0$, photon, $Z^0$--photon self-energy diagrams are shown
in~Figs.~\ref{fig:NMSSM_W}, \ref{fig:NMSSM_Z},
\ref{fig:NMSSM_g}, \ref{fig:NMSSM_Zg}, respectively.



\begin{thebibliography}{99}


\bibitem{Peskin:1990zt}
  M.~E.~Peskin, T.~Takeuchi,
  ``A New constraint on a strongly interacting Higgs sector,''
  \mbox{Phys.\ Rev.\ Lett.\  {\bf 65}, 964-967 (1990).}

\bibitem{Kennedy:1990ib} 
  D.~C.~Kennedy and P.~Langacker,
  ``Precision electroweak experiments and heavy physics: A Global analysis,''
  \mbox{Phys.\ Rev.\ Lett.\  {\bf 65}, 2967 (1990)}
  \mbox{[Erratum-ibid.\  {\bf 66}, 395 (1991)].}
  
\bibitem{Peskin:1991sw} 
  M.~E.~Peskin and T.~Takeuchi,
  ``Estimation of oblique electroweak corrections,''
  \mbox{Phys.\ Rev.\ D {\bf 46}, 381 (1992).}
  
\bibitem{Maksymyk:1993zm}
  I.~Maksymyk, C.~P.~Burgess and D.~London,
  ``Beyond S, T and U,''
  \mbox{Phys.\ Rev.\ D {\bf 50}, 529 (1994)}
  [hep-ph/9306267].

\bibitem{Burgess:1993mg}
  C.~P.~Burgess, S.~Godfrey, H.~Konig, D.~London, I.~Maksymyk,
  ``A Global fit to extended oblique parameters,''
  \mbox{Phys.\ Lett.\  {\bf B326}, 276-281 (1994)}
  [hep-ph/9307337].


\bibitem{Nakamura:2010zzi}
  K.~Nakamura {\it et al.} [Particle Data Group Collaboration],
  ``Review of particle physics,''
  \mbox{J.\ Phys.\ G {\bf G37}, 075021 (2010).}

\bibitem{Maniatis:2009re}
  M.~Maniatis,
  ``The Next-to-Minimal Supersymmetric extension of the Standard Model reviewed,''
  Int.\ J.\ Mod.\ Phys.\  {\bf A25}, 3505-3602 (2010).
  [arXiv:0906.0777 [hep-ph]].

\bibitem{Ellwanger:2009dp}
  U.~Ellwanger, C.~Hugonie, A.~M.~Teixeira,
  ``The Next-to-Minimal Supersymmetric Standard Model,''
  Phys.\ Rept.\  {\bf 496}, 1-77 (2010).
  [arXiv:0910.1785 [hep-ph]].



\bibitem{2001yb}
  [LEP Higgs Working Group for Higgs boson searches Collaboration],
  ``Flavor independent search for hadronically decaying neutral Higgs bosons at LEP,''
  \mbox{[hep-ex/0107034].}


\bibitem{Domingo:2011uf} 
  F.~Domingo and T.~Lenz,
  ``W mass and Leptonic Z-decays in the NMSSM,''
  \mbox{JHEP {\bf 1107}, 101 (2011)}
  \mbox{[arXiv:1101.4758 [hep-ph]].}
  

\bibitem{Dedes:1998hg}
  A.~Dedes, A.~B.~Lahanas, K.~Tamvakis,
  ``The Effective weak mixing angle in the MSSM,''
  \mbox{Phys.\ Rev.\  {\bf D59}, 015019 (1999)}
  \mbox{[hep-ph/9801425].}

\bibitem{Heinemeyer:2004gx}
  S.~Heinemeyer, W.~Hollik, G.~Weiglein,
  ``Electroweak precision observables in the minimal supersymmetric standard model,''
  \mbox{Phys.\ Rept.\  {\bf 425}, 265-368 (2006)}
  \mbox{[hep-ph/0412214].}

\bibitem{RamseyMusolf:2006vr}
  M.~J.~Ramsey-Musolf, S.~Su,
  ``Low Energy Precision Test of Supersymmetry,''
  \mbox{Phys.\ Rept.\  {\bf 456}, 1-88 (2008)}
  \mbox{[hep-ph/0612057].}
  
\bibitem{Cho:2011rk}
  G.~-C.~Cho, K.~Hagiwara, Y.~Matsumoto, D.~Nomura,
  ``The MSSM confronts the precision electroweak data and the muon g-2,''
  \mbox{[1104.1769 [hep-ph]].}

\bibitem{Gunion:2012zd}
  J.~F.~Gunion, Y.~Jiang and S.~Kraml,
  ``The Constrained NMSSM and Higgs near 125 GeV,''
  \mbox{Phys.\ Lett.\ B {\bf 710} (2012) 454}
  \mbox{[arXiv:1201.0982 [hep-ph]].}
    
\bibitem{Djouadi:2008uw} 
  A.~Djouadi, M.~Drees, U.~Ellwanger, R.~Godbole, C.~Hugonie, S.~F.~King, S.~Lehti and S.~Moretti {\it et al.},
  ``Benchmark scenarios for the NMSSM,''
  \mbox{JHEP {\bf 0807}, 002 (2008)}
  \mbox{[arXiv:0801.4321 [hep-ph]].}
  
\bibitem{Ellwanger:2006rn}
  U.~Ellwanger, C.~Hugonie,
  ``NMSPEC: A Fortran code for the sparticle and Higgs masses in the NMSSM with GUT scale boundary conditions,''
  \mbox{Comput.\ Phys.\ Commun.\  {\bf 177}, 399-407 (2007)}
  \mbox{[hep-ph/0612134].}

\bibitem{King:2012is} 
  S.~F.~King, M.~Muhlleitner and R.~Nevzorov,
  ``NMSSM Higgs Benchmarks Near 125 GeV,''
  \mbox{Nucl.\ Phys.\ B {\bf 860}, 207 (2012)}
  \mbox{[arXiv:1201.2671 [hep-ph]].}

\bibitem{AbdusSalam:2011fc} 
  S.~S.~AbdusSalam, B.~C.~Allanach, H.~K.~Dreiner, J.~Ellis, U.~Ellwanger, J.~Gunion, S.~Heinemeyer and M.~Kraemer {\it et al.},
  ``Benchmark Models, Planes, Lines and Points for Future SUSY Searches at the LHC,''
  \mbox{Eur.\ Phys.\ J.\ C {\bf 71}, 1835 (2011)}
  \mbox{[arXiv:1109.3859 [hep-ph]].}


\bibitem{Christensen:2008py}
  N.~D.~Christensen, C.~Duhr,
  ``FeynRules - Feynman rules made easy,''
  \mbox{Comput.\ Phys.\ Commun.\  {\bf 180}, 1614-1641 (2009)}
  \mbox{[arXiv:0806.4194 [hep-ph]].}

\bibitem{Duhr:2011se}
  C.~Duhr and B.~Fuks,
  ``A superspace module for the FeynRules package,''
  \mbox{Comput.\ Phys.\ Commun.\  {\bf 182}, 2404 (2011)}
  [arXiv:1102.4191 [hep-ph]].
  
\bibitem{arXiv:0801.0045} 
  B.~C.~Allanach, C.~Balazs, G.~Belanger, M.~Bernhardt, F.~Boudjema, D.~Choudhury, K.~Desch and U.~Ellwanger {\it et al.},
  ``SUSY Les Houches Accord 2,''
  \mbox{Comput.\ Phys.\ Commun.\ \ {\bf 180}, 8  (2009)}
  \mbox{[arXiv:0801.0045 [hep-ph]].}

\bibitem{Hahn:2000kx}
  T.~Hahn,
  ``Generating Feynman diagrams and amplitudes with FeynArts 3,''
  \mbox{Comput.\ Phys.\ Commun.\  {\bf 140}, 418-431 (2001)}
  \mbox{[hep-ph/0012260].}

\bibitem{Hahn:1998yk}
  T.~Hahn, M.~Perez-Victoria,
  ``Automatized one loop calculations in four-dimensions and D-dimensions,''
  \mbox{Comput.\ Phys.\ Commun.\  {\bf 118}, 153-165 (1999)}
  \mbox{[hep-ph/9807565].}

\bibitem{ccode}
C-code for the oblique
parameters is available at the url~\url{www.physik.uni-bielefeld.de/~maniatis/NMSSM/oblique}.
  
\end{thebibliography}
\end{document}